# Single-Crystalline Metallic Films Induced by van der Waals Epitaxy on Black Phosphorus

*Yangjin Lee,[1,2] Han-gyu Kim,[1] Tae Keun Yun,[1] Jong Chan Kim,[3] Sol Lee,[1,2] Sung Jin Yang,[1] Myeongjin Jang,[1,2] Donggyu Kim,[1] Huije Ryu,[4] Gwan-Hyoung Lee,[4,5] Seongil Im,[1] Hu Young Jeong,[3,6,*] Hyoung Joon Choi,[1,*] and Kwanpyo Kim[1,2,*]*

[1]Department of Physics, Yonsei University, Seoul 03722, South Korea

[2]Center for Nanomedicine, Institute for Basic Science, Seoul 03722, South Korea

[3]Department of Materials Science and Engineering, Ulsan National Institute of Science and Technology, Ulsan 44919, South Korea

[4]Materials Science and Engineering, Seoul National University, Seoul 08826, South Korea

[5]Research Institute of Advanced Materials (RIAM), Institute of Engineering Research, Institute of Applied Physics, Seoul National University, Seoul 08826, South Korea

[6]UNIST Central Research Facilities, Ulsan National Institute of Science and Technology, Ulsan 44919, South Korea

* Address correspondence to K.K. (kpkim@yonsei.ac.kr), H.J.C. (h.j.choi@yonsei.ac.kr), and H.Y.J. (hulex@unist.ac.kr)

**Abstract:** The properties of metal-semiconductor junctions are often unpredictable because of non-ideal interfacial structures, such as interfacial defects or chemical reactions introduced at junctions. Black phosphorus (BP), an elemental two-dimensional (2D) semiconducting crystal, possesses the puckered atomic structure with high chemical reactivity, and the establishment of a realistic atomic-scale picture of BP's interface toward metallic contact has remained elusive. Here we examine the interfacial structures and properties of physically-deposited metals of various kinds on BP. We find that Au, Ag, and Bi form single-crystalline films with (110) orientation through guided van der Waals epitaxy. Transmission electron microscopy and X-ray photoelectron spectroscopy confirm that atomically sharp van der Waals metal-BP interfaces forms with exceptional rotational alignment. Under a weak metal-BP interaction regime, the BP's puckered structure play an essential role in the adatom assembly process and can lead to the formation of a single crystal, which is supported by our theoretical analysis and calculations. The experimental survey also demonstrates that the BP-metal junctions can exhibit various types of interfacial structures depending on metals, such as the formation of polycrystalline microstructure or metal phosphides. This study provides a guideline for obtaining a realistic view on metal-2D semiconductor interfacial structures, especially for atomically puckered 2D crystals.

## Introduction

Metal-semiconductor junctions are an indispensable part of electronic devices and so have been the subject of a great deal of research activities in the past several decades[1, 2]. Various aspects of these interfaces, such as the atomic configuration of metal-semiconductor interfaces, formation mechanisms of the Schottky barrier, junction-based device operation, and the reduction of contact resistance, have been studied[1, 2]. Recently, one-dimensional (1D) and two-dimensional (2D) semiconducting crystals have shown the potential for use in next-generation electronics applications[3-6]. Even though the study of metal-semiconducting junction for these nanocrystals is of great importance, the unconventional contact geometry to nanocrystals and the degraded quality of interface from device fabrication process make it quite challenging to properly address important junction-related issues[7-14].

Black phosphorus (BP), a layered elemental van der Waals 2D crystal, has recently received a tremendous amount research attention because of its versatile electrical properties, which include a tunable direct bandgap, high charge carrier mobility, and in-plane anisotropic properties[15-20]. However, the quality of BP crystal can be easily undermined by uncontrolled chemical processes on its surface, for example by exposure to ambient environment or by surface damage induced during device fabrication[21-23]. A significant amount of research has been devoted to passivate BP surfaces via mainly through chemical reactions, which can form protective layers on BP and increase the reliability of device operation[24-29]. Although the previous inactivation of BP surface can be utilized for achieving stable BP-dielectric interface, BP-metal interfaces have been relatively under-investigated[30-33]. Moreover, an unusual van der Waals epitaxy process may occur during the assembly of metal atoms on the BP's surface due to its puckered structure[34-36]. This process may lead to the formation of assembly structure, which is different from those obtained on other hexagonal 2D materials, such graphene and transition metal dichalcogenides (TMDCs)[11, 37-40].

Here we systematically studied the growth behavior of various metallic films on BP and investigate their properties. We found that the single-crystalline Au, Ag, and Bi films form on BP through guided assembly along the BP's atomic trenches. In-depth transmission electron microscopy (TEM) and X-ray photoelectron spectroscopy (XPS) characterizations were performed to investigate the microstructures and properties of metal-BP junctions, which clearly shows the formation of atomically sharp metal interface on BP surface with exceptional rotational alignment (Metal [110] // BP [010] and metal [110] // zigzag direction of BP). The anisotropic BP template allows for the growth of single-crystalline films even by means of physical deposition at room temperature. The guided van der Waals assembly of metal atoms and the formation of single-crystalline films can be achieved under a weak metal-BP interaction regime, which was supported by theoretical calculations and analysis. The different interfacial structures, such as the formation of polycrystalline film or metal phosphide at interface, were observed with metals possessing the stronger BP-metal interaction. The observed wide range of possible metal-BP interfacial structures provide insight on establishing a realistic picture on interfacial structure between metal and 2D semiconducting crystals, especially atomically puckered 2D crystals with high chemical reactivity.

## Results and Discussion

A schematic of how a single-crystalline metallic film grows on a BP template is shown in Figure 1a. After metal atoms are adsorbed on the BP's surface via physical deposition, metal adatoms are preferentially aligned along the BP's atomic trenches due to the highly anisotropic surface diffusion of atoms and energetically preferred 1D assembly configuration. As neighboring aligned metallic chains merge, they produce a well-oriented metallic film. In this study, various kinds of metals were deposited onto BP substrates using conventional physical deposition techniques. We tested various metal deposition processes and found that the microstructure of metallic films on BP depend on the deposition methods and conditions. We also found that guided van der Waals epitaxy can induce the formation of single-crystalline metallic films for metallic species showing weak interaction toward BP.

The epitaxy behavior of Au on BP can be regarded as an exemplary and the results on Au-BP are summarized in Figure 1b-h. The microstructure and relative crystal orientation of Au on BP were characterized in various ways using TEM. We took caution to maintain the pristine surface quality of BP during the sample preparation (See Method and Supporting Figure S1). The selected area electron diffraction (SAED) of plan-view samples shows the crystallographic orientation of Au relative to the BP (Figure 1b). SAED of the Au/BP heterostructure showed two sets of diffraction spots, one set from BP diffraction with [010] zone axis and the other set from the Au [110] pattern. The inter-planar spacing of $(002)_{Au}$ and $(220)_{Au}$ were 2.04 Å and 1.44 Å, respectively, which is consistent with that of bulk Au. The crystal direction $(220)_{Au}$ was aligned with one zigzag direction of the BP while the other lattice plane orientation of $(220)_{Au}$ was parallel to the out-of-plane direction of the BP. Therefore, the crystallographic relationship at heteroepitaxial interface of Au and BP can be characterized as $(002)_{Au}//(002)_{BP}$, $(\bar{2}20)_{Au}//(200)_{BP}$, $[110]_{Au}//[010]_{BP}$.

The crystallinity and the degree of rotational alignment of Au films on BP depends on the deposition methods (e-beam evaporation, thermal evaporation, and DC sputtering; Figure 1c and Supporting Figure S2, S3). E-beam evaporation produced the best-aligned Au film growth. Thermal evaporation produced a slightly lower degree of crystallinity than e-beam evaporation. Moreover, a different orientation of Au film, [111] Au, was also observed, resulting in the formation of Au films with mixture of [110] and [111] (Supporting Figure S2a). Nevertheless, the Au [111] diffraction pattern still maintained in-plane rotational alignment with respect to the BP substrate. On the other hand, DC sputtering resulted in the formation of polycrystalline Au film (Supporting Figure S2b). Figure 1c shows the azimuthal intensity plot of BP (200) peak and Au ($\bar{2}20$) peak produced by different deposition methods. The FWHM was 1.6° for e-beam deposition, 2.3° for thermal deposition, and 12.6° for DC sputtering.

The Au grows on the BP by following typical Volmer-Weber growth[41] as shown in a TEM image (Figure 1d), where Au nanoparticles first form in an early stage (0.5 nm thick) of film growth. The Au nanoparticles displayed moiré fringes produced by the lattice mismatch between the Au and the BP. Figure 1e shows an image of the Au/BP moiré pattern which had a uniform periodicity of 1.14 nm along the BP zigzag lattice direction. The measured distance is consistent with the moiré distance calculated from the Au/BP model (Figure 1f). The uniform moiré pattern also confirms that the grown Au structure is highly aligned due to the BP growth template.

The BP-templated assembly of Au on BP enables the formation of large-area single-crystalline Au film. To verify this, electron back scattering diffraction (EBSD) mapping was used on 50 nm-thick Au films deposited on BP (Figure 1g). It is clear that the single-crystalline Au films forms on a BP flake when the Au was deposited by e-beam evaporation. On the other hand, Au film deposited by DC sputtering did not exhibit the preferred alignment behavior on BP as shown in Figure 1h. Au films prepared on BP showed a low degree of roughness as confirmed by AFM imaging (Supporting Figure S4).

The interface quality and microstructure of Au/BP heterostructures were investigated in detail by cross-section STEM imaging as shown in Figure 2. For this goal, multiple cross-section samples along BP [100] and [001] zone axes were prepared. Figure 2a and Supporting Figure S5 show annular bright field (ABF)-STEM images at BP [100] zone axis. The formation of single-crystalline Au film was clearly visualized by STEM images and their Fast Fourier Transform (FFT) signals (Figure 2b and Supporting Figure S5b). We found the atomically sharp interface with 1.9 Å Au-BP interatomic distance along the vertical direction and no hint of an additional wetting layer (amorphous phase) as shown in Figure 2c. The STEM imaging at a different zone axis (BP [001] zone axis) also demonstrated the formation of high-quality Au film with atomically sharp interface toward BP (Figure 2e-g and Supporting Figure S6), which is consistent with other STEM data. The epitaxial formation of single-crystalline Au films was also maintained on few-layer BP samples, which indicates that the observed epitaxy process does not depend on the thickness of BP substrate (Supporting Figure S7).

We note that the report of single-crystalline metallic films grown on 2D crystals via physical deposition at room temperature has been very rare. For example, Au exhibited a typical polycrystalline film formation on a graphene (Supporting Figure S8)[42]. Au crystallinity can be improved by conducting metal deposition at elevated substrate temperatures or by conducting post-annealing processes[42, 43]. Similarly, the crystallinity of Au deposited at room temperature on TMDCs was limited[44, 45]. Moreover, Au on a BP substrate exhibited (110) interface, not the more common (111) Au interface on other 2D crystals, including graphene and TMDCs. Our calculation results indicated that the binding energy for Au (110) interface on BP is comparable to that of Au (111) interface on BP; the interfacial binding energies for Au(110)/BP and Au(111)/BP are 1.312 eV and 1.310 eV per BP unit cell area, respectively (Supporting Figure S9). The extra calculation results with sub-monolayer coverage of Au on BP showed that (110) orientation was preferred at the initial stage of film formation. This is likely due to the puckered topography of (110) surface and its stronger interaction with BP (Supporting Figure S10). This calculation results are also consistent with the relatively small lattice mismatch (< 8%) of (110) metal films and BP lattice along the puckered direction (AC direction) as shown in Supporting Table S1.

BP surface quality plays has a strong influence over the results of van der Waals epitaxy of metals. The surface of BP is prone to degradation from exposure to the ambient environment or during sample fabrication, so it is vital to experimentally verify BP surface quality. From cross-section TEM imaging (Supporting Figure S11), we found that DC sputtering, which generally degrades the template surface more than e-beam evaporation[46], resulted in disordering the top regions (~1.1 nm) of BP. The formation of misaligned Au grains by DC sputtering can be attributed to the observed disordered BP interface. Similarly, when the BP was exposed to ambient environment prior to Au deposition, the deposited Au film by e-beam evaporation also showed lower crystallinity and increased misalignment with increasing exposure duration (Supporting Figure S12).

The formation of single-crystalline metallic films was observed using different metallic species, Ag and Bi. Figure 3a shows the SAED of single-crystalline Ag/BP, which showed two sets of diffraction spots similar to Au/BP. The relative crystal orientation of Ag film on BP was identical to that of Au/BP: $(002)_{Ag}//(002)_{BP}$, $(\bar{2}20)_{Ag}//(200)_{BP}$, $[110]_{Ag}//[010]_{BP}$. The small FWHM of the $(\bar{2}20)_{Ag}$ peak (~ 0.9°) verifies a highly crystalline structure and nearly perfect rotational alignment (Figure 3b). The uniform moiré patterns with 1.14 nm periodicity were observed, which were also identical to those in the Au/BP heterostructure (Figure 3c).

Bi also showed remarkable epitaxial behavior on the BP assembly template. Figure 3d shows the SAED pattern of the Bi/BP heterostructure which clearly shows a single-crystalline Bi formation on BP. Electron diffraction pattern of Bi shows that the rectangular unit cell of Bi formed with the lattice constants of 4.54 Å and 4.75 Å, which is in agreement with the pseudo-cubic Bi

(110) lattice constant[47]. The FWHM of $(112)_{Bi}$ was ~0.9 ° (Figure 3e), indicating that the Bi film deposited on the BP also has a high level of crystallinity. Atomic-resolution STEM image of the Bi-BP interface shows the atomically sharp interface without apparent damage on BP (Figure 3f and 3g). Interestingly, the Bi film changed its crystalline orientation from $(110)_{Bi}//(010)_{BP}$ to $(111)_{Bi}//(010)_{BP}$ when the thickness of the Bi film was greater than 4 nm (Supporting Figure S13). Nevertheless, the Bi/BP interface maintained the locked-in relative epitaxial relation. The previous studies[48, 49] have reported the similar switch of Bi film orientation at the similar Bi thickness, which was attributed to the competition between the surface and bulk energies in (110)- or (111)-terminated Bi films.

The physical deposition of other species of metals on BP exhibits different microstructure formation. Based on our observation, we can categorize the types of interfacial microstructures into three categories (Figure 4a): the formation of previously discussed single-crystalline films ("type-1"), poly-crystalline films without metal phosphide formation ("type-2"), and poly-crystalline films with the formation metal phosphides ("type-3"). Figures 4b, 3c and Supporting Figure S14 show SAED examples of type-2 and type-3 microstructures. When Pt, Al, Ti, and Cr were deposited on BP, they showed polycrystalline microstructures (type-2). Meanwhile, when Cu and In were used, polycrystalline microstructures with the formation of metal phosphides were observed (type-3). AFM characterization of type-3 metal films on BP revealed a large surface roughness due to the non-uniform formation of metal phosphides (Supporting Figure S15). To understand the formation of metal phosphides even by physical deposition for some metals, the Gibbs free energies for the formation of metal phosphides were analyzed (Supporting Table S2). We found that the metal species that formed single-crystalline films had positive Gibbs free energies, indicating that the formation of metal phosphides is not energetically favorable at room temperature. All other metals had negative Gibbs free energies for metal phosphide formation, which is consistent with the observed formation of phosphides by metal deposition at room temperature.

XPS was conducted to investigate the metal-BP interfacial properties in detail. Figure 4d shows the XPS spectra of the P 2p core levels for bare BP and metal-BP junctions. In the upper panel of Figure 4d, the P core level from bare BP samples shows the typical characteristic doublet of P $2p_{3/2}$ and P $2p_{1/2}$ with spin-orbit splitting of approximately 0.9 eV. The peak positions of the doublet is consistent with previously reported results of 130.06 and 130.96 eV, respectively[27]. The XPS spectra of the P 2p core level upon metal deposition revealed that the overall shapes of the P $2p_{3/2}$ and P $2p_{1/2}$ peaks were unaffected for metals that formed type-1 microstructures (Au, Ag, and Bi). Peak position analyses indicated that the peaks shifted by ~0.1 eV to a lower binding energy for type-1 metals (Figure 4e). The P binding level shift can be explained by the energy band diagram of the metal-BP interface (Figure 4f). Due to the difference in work functions of BP (~ 4.2 eV) and the metals (> 4.2 eV), electrons were transferred from BP to the metals upon junction formation, causing upward band bending for BP [50]. The degree of band bending should be smaller than the bandgap of bulk BP (~ 0.4 eV), which is consistent with the observed binding level shifts. The P 2p core level from DC sputtered Au on BP showed the down-shift (~0.1 eV) compared to that of single crystalline Au on BP (Supporting Figure S16). The damaged BP surface during DC sputtering process may serve as an intermediate charge transfer layer, which results in the modification in the degree of charge transfer and band bending at BP.

The XPS spectra also confirmed that the deposition of Pt (type-2) and Cu (type-3) produced strong changes in BP surface (Figs. 4d and 4e). Pt-BP samples showed a peak shift of relatively large value of ~ 0.8 eV, indicating that Pt-phosphorus interaction is quite strong.

Moreover, XPS showed that the peak shape of Cu-BP changed after deposition. The deconvolution of peaks indicates the formation of copper phosphide ($Cu_3P$)[51], which was consistent with the TEM and SAED results.

To understand the formation of single-crystal films and other types of microstructure from metal deposition on BP, first-principles calculations were performed. For calculations, we changed the position of an individual metal adatom relative to BP substrate (Figure 5a) and calculated the equilibrium height and adsorption energies. This process generated a 2D adsorption energy landscape of metal adatom on BP. The exemplary calculation results with Au (Figure 5b) and Cu (Figure 5c) show that the favorable adsorption sites display 1D linear patterns along the BP zigzag lattice direction. The highly anisotropic adsorption energies are also evident from the energy barriers to adatom diffusion in different directions (Figure 5d). The energy barrier along the zigzag lattice direction is low, essentially zero for Au, which is in agreement with the results of other studies[34, 35, 52, 53]. This result supports our hypothesis that adatoms form 1D chains along the BP zigzag direction and that inter-chain assemblies induce the formation of single-crystalline film with the underlying BP (Figure 1a). The (110) metallic surface is also consistent with our picture that the epitaxy is driven by inter-chain assembly process as shown in Figure 1a.

The calculation results for the type-2 and type-3 metals were also consistent with the experimental observations. Type-2 metals exhibited significantly larger adsorption energies than the type-1 and type-3 metals as shown in Figure 5e, indicating a stronger metal-BP interaction. The large adsorption energies of type-2 metals suggest that the type-2 metals are strongly anchored on the surface of BP with limited diffusion and reactivity on BP, resulting in the formation of polycrystalline films. The calculation results for Cu (type-3) indicated that it had adsorption energy and diffusion barrier values similar to those of type-1 metals. However, the adsorption energy landscape of Cu showed that the energetically preferred adsorption sites were arranged in meandering patterns (Figure 5c). The 1D van der Waals assembly process will be ineffective under this adsorption potential landscape and the recently observed formation of chiral Cu structure on BP could be related to our calculation results[36]. We caution that the proper treatment on the formation of metal phosphide should take account the role of chemical reaction between metal and phosphorus, rather than interaction between single adatom and BP.

## Conclusions

This study demonstrated that BP's atomic trenches can serve as efficient van der Waals templates for forming single-crystalline metal films. Extensive structural characterizations showed that single-crystalline Au, Ag, and Bi metal films were successfully grown on BP using epitaxy. The epitaxial alignment of metals on BP was found to be driven by the highly anisotropic diffusion and energy characteristics of the metal atoms on BP, which was supported by theoretical calculations. This anisotropic assembly process formed energetically meta-stable (110) metallic film interfaces on BP. The observed wide range of possible interfacial structures at metal-BP provides insight on establishing a realistic picture on metal-2D semiconductor interfaces, especially for atomically puckered 2D crystals, such as IV-VI monochalcogenides.

## Methods

**Sample Preparation.** BP flakes (Smart Elements) were mechanically exfoliated onto a PDMS substrate and subsequently transferred to $Si_3N_4$ TEM grids or $SiO_2$/Si wafers. BP samples were prepared using methods described in previous studies (See Supporting Figure S1)[54, 55]. To minimize BP surface degradation, the entire sample fabrication process was performed inside a nitrogen-filled glove box with an oxygen concentration of less than 0.1 ppm except for the sample loading process during metal deposition. Thin BP flakes were identified using an optical microscope (Leica DM 750M) under transmission mode. Au was deposited on BP using thermal evaporation, e-beam evaporation, and DC magneton sputtering. Ag, Cu and Cr were deposited on BP using thermal evaporation and e-beam evaporation. Bi, Ti and In were deposited on BP using thermal evaporation. Pt and Al were deposited on BP using e-beam evaporation. The target sample was kept at room temperature for the duration of metal deposition processes. Deposition occurred at a rate of 0.1 Å/s and the vacuum level was kept below $2.0 \times 10^{-6}$ Torr (thermal evaporation) and $3.0 \times 10^{-7}$ Torr (e-beam evaporation). DC magnetron sputtering was performed with a sputtering power of 10 W under a background pressure of less than $1.1 \times 10^{-6}$ Torr. TEM cross-section samples were fabricated using a focused ion beam process (Helios NanoLab 450, FEI). Metal deposition of thicknesses of 2 or 4 nm was used for XPS characterizations.

**Characterizations.** TEM images and diffraction patterns were acquired using FEI Tecnai F-20 and JEOL-2010Plus operated at 200kV. HR-TEM and HR-STEM images were acquired with double Cs-aberration corrected FEI Titan G2 and JEOL ARM-200F, which operated at 80kV and 200 kV. EBSD scanning was performed using an FEI Quanta 3D FEG with an EDAX-TSL EBSD system. AFM measurement was performed with a Park Systems XE-7 under ambient conditions. XPS analysis was conducted on a Thermo Fisher K-alpha spectrometer with 100-μm beam size.

**First-principles Calculations.** We performed density functional theory (DFT) calculation with the SIESTA[56] code, using the Perdew−Burke−Ernzerhof-type generalized gradient approximation (PBE-GGA)[57] for the exchange-correlation energy, fully relativistic norm-conserving pseudopotentials[58], pseudoatomic orbitals of the double-zeta polarization basis set, a uniform real-space mesh generated with the cutoff energy of 500 Ry, and the DFT-D2 scheme[59] for the van der Waals interaction. We determined equilibrium lattice constants of monolayer black phosphorus (BP) by minimizing the total energy, using a 16×16 k-point grid for integration of the density over the Brillouin zone (BZ). Obtained equilibrium lattice constants are $a$ = 3.3597 Å and $b$ = 4.5361 Å. To simulate metal-atom adsorption, we used a 4×3 supercell of monolayer BP, where 4 is along the zigzag direction and 3 along the armchair direction, and added a single metal atom in the supercell. To obtain the binding energy of the adatom as a function of its lateral position, we relaxed positions of all phosphorus atoms and the height of the adatom, using a 3×3 k-point grid in the supercell BZ and including the spin-orbit coupling. The adsorption energy of an adatom is defined as $E_{ads} = E_{ad} - E_{BP} - E_{atom}$, where $E_{ad}$ is the total energy of monolayer BP with an adatom on it, $E_{BP}$ is the total energy of pristine monolayer BP, and $E_{atom}$ is the total energy of an isolated metal atom.

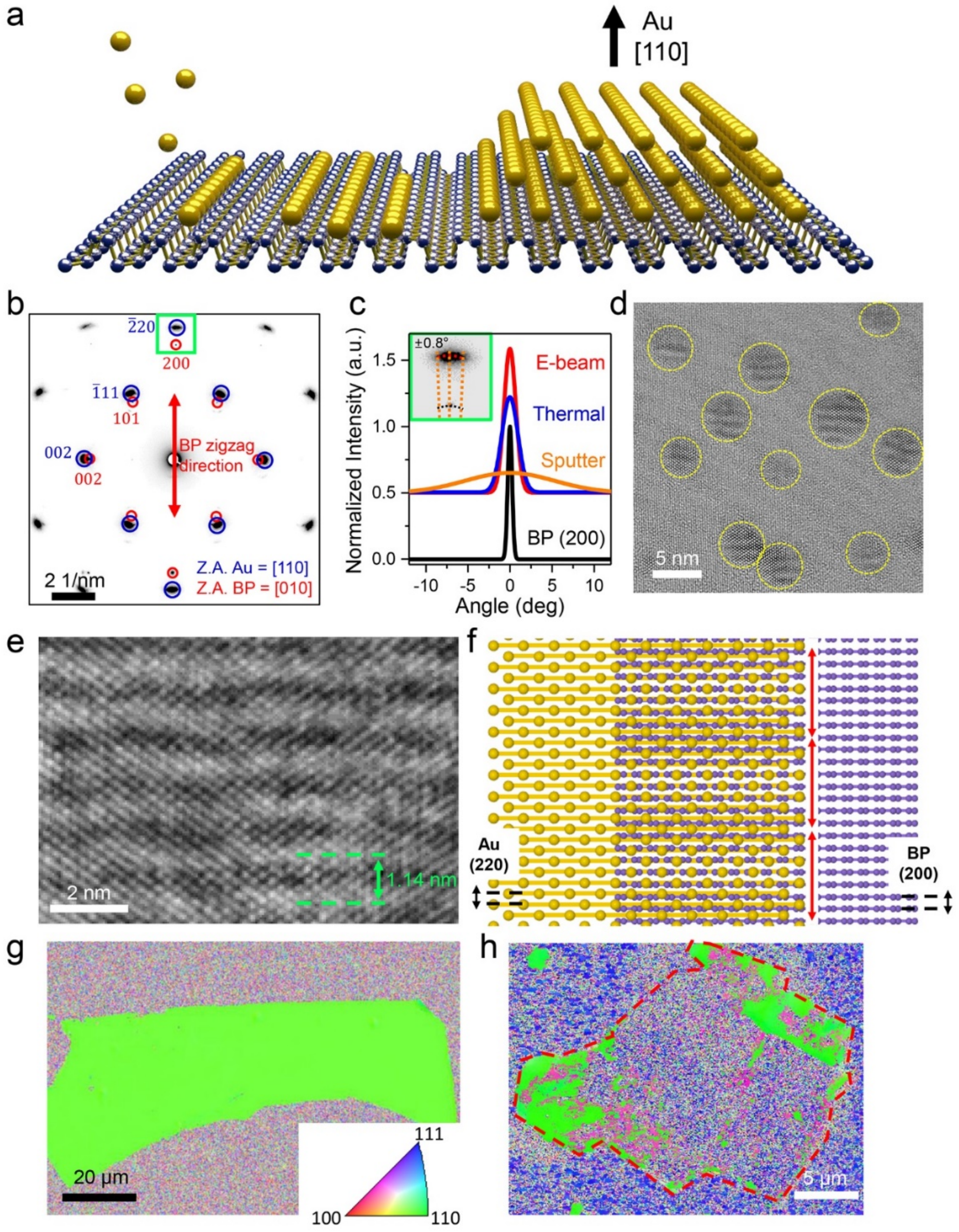

**Figure 1. Van der Waals epitaxy of single-crystalline Au film on BP.** (a) Schematic illustration of metal atom adsorption and film formation process on BP. (b) SAED pattern of Au/BP heterostructure. Diffraction spots from Au and BP are marked by blue and red circles, respectively. (c) Azimuthal intensity plot of BP (200) peak and Au ($\bar{2}20$) peak. Inset: enlarged diffraction signal denoted by the green box in panel b. (d) TEM image of Au nanoparticles grown on BP under the low Au coverage condition. (e) HR-TEM image and (f) atomic model of Au film (110 top surface) on BP projected along the BP [010] axis. Moiré fringe with 1.14 nm periodicity indicated. (g) EBSD mapping of single crystalline Au on a BP flake formed by e-beam evaporation. (h) EBSD mapping of Au on a BP flake formed by DC Sputtering.

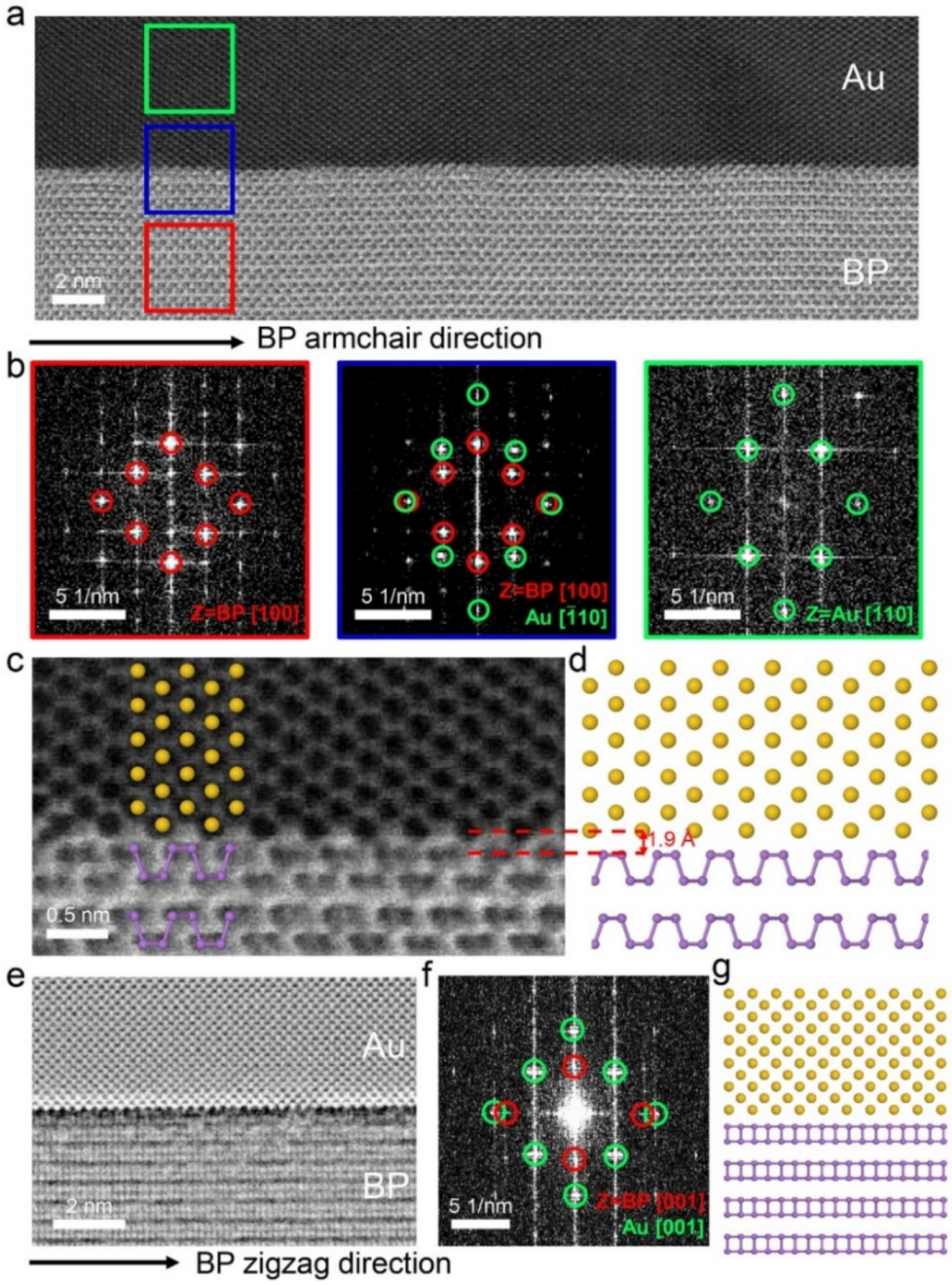

**Figure 2. Cross-sectional STEM characterization of single crystalline Au on BP.** (a) ABF-STEM image of the single crystalline Au on BP. Green, blue, red colored box indicates the Au, Au/BP interface, and BP regions, respectively. (b) FFT signals from the three areas indicated in panel (a). Au and BP diffraction peaks are marked with green and red colored dots, respectively. (c) Atomic resolution ABF-STEM image and (d) atomic model of the Au/BP interface projected along the BP [100] axis. (e) HAADF-STEM image, (f) FFT signal, and (g) atomic model of the Au/BP interface projected along the BP [001] axis. A band-pass filter was used to enhance the contrast of this image.

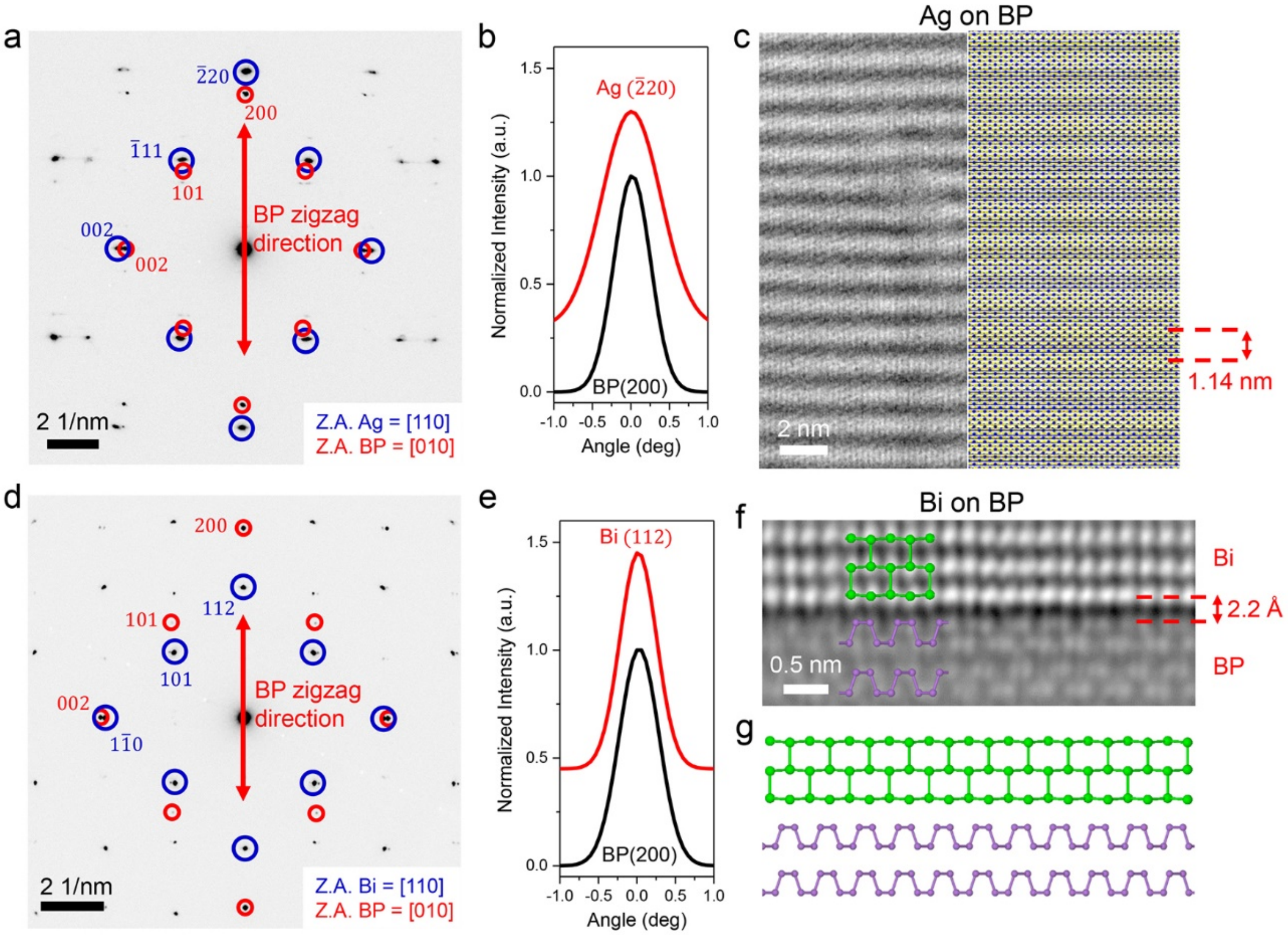

**Figure 3. Epitaxial growth of Ag and Bi films on BP.** (a) SAED pattern of Ag/BP heterostructure. (b) Azimuthal intensity plot of BP (200) peak and Ag ($\bar{2}20$) peak. (c) High-magnification TEM micrograph (left) and atomic model of Ag (110) on BP (right). (d) SAED pattern of Bi/BP heterostructure. (e) Azimuthal intensity plot of BP (200) peak and Bi (112) peak. (f) Atomic resolution HAADF-STEM image and (g) atomic model of Bi/BP interface projected along the BP [100] axis. A band-pass filter was used to enhance the contrast of this image.

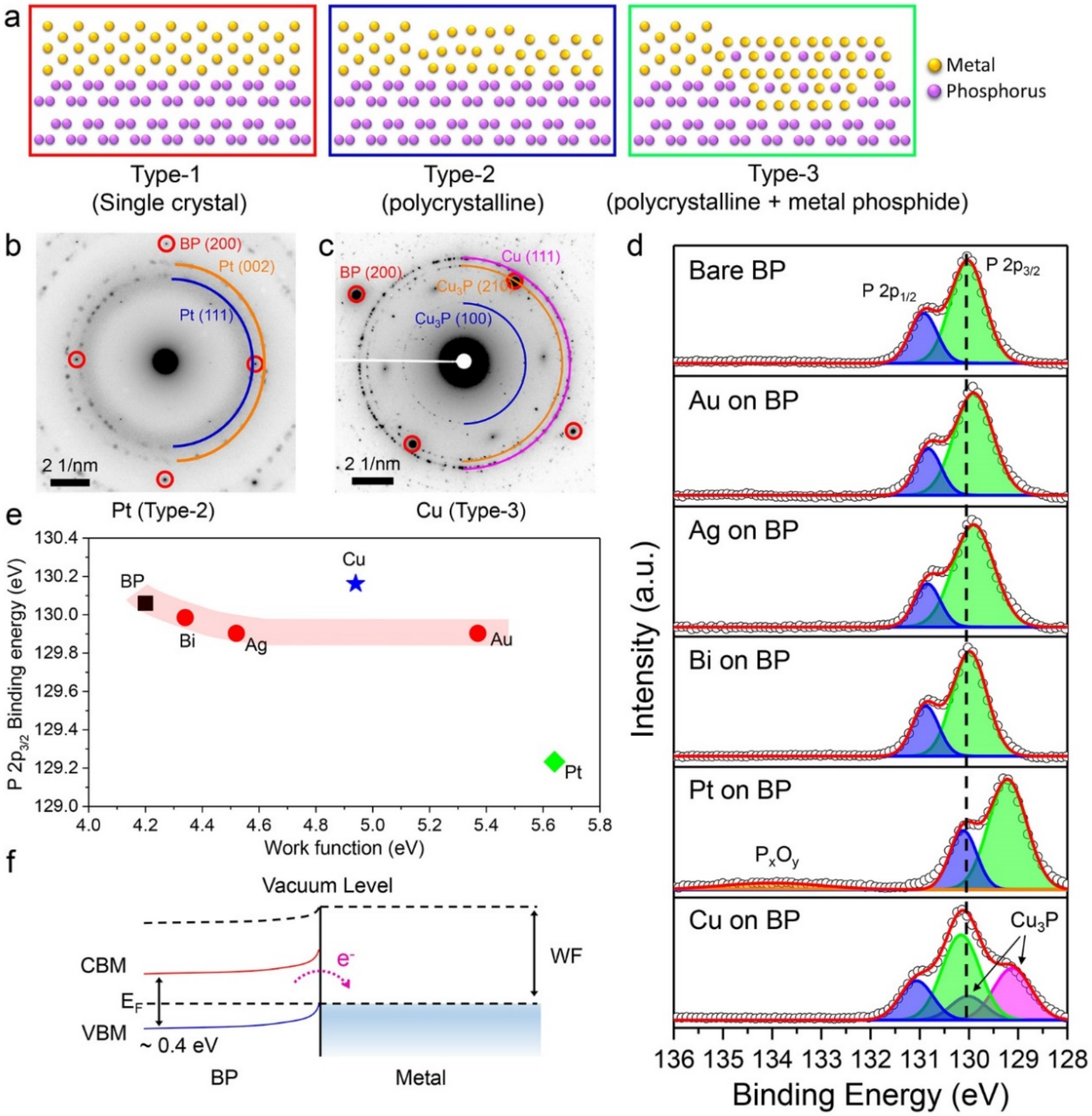

**Figure 4. X-ray photoelectron spectroscopy of the metal-BP interface.** (a) Metal categories showing different types of interfacial microstructures. SAED of (b) 5 nm-thick Pt and (c) 5 nm-thick Cu deposited on BP. (d) XPS of the P 2p core levels of the metal-BP interface. (e) Plot of P $2p_{2/3}$ binding energy with different metal contacts. (f) Schematic illustration of the band diagram of the metal-BP interface.

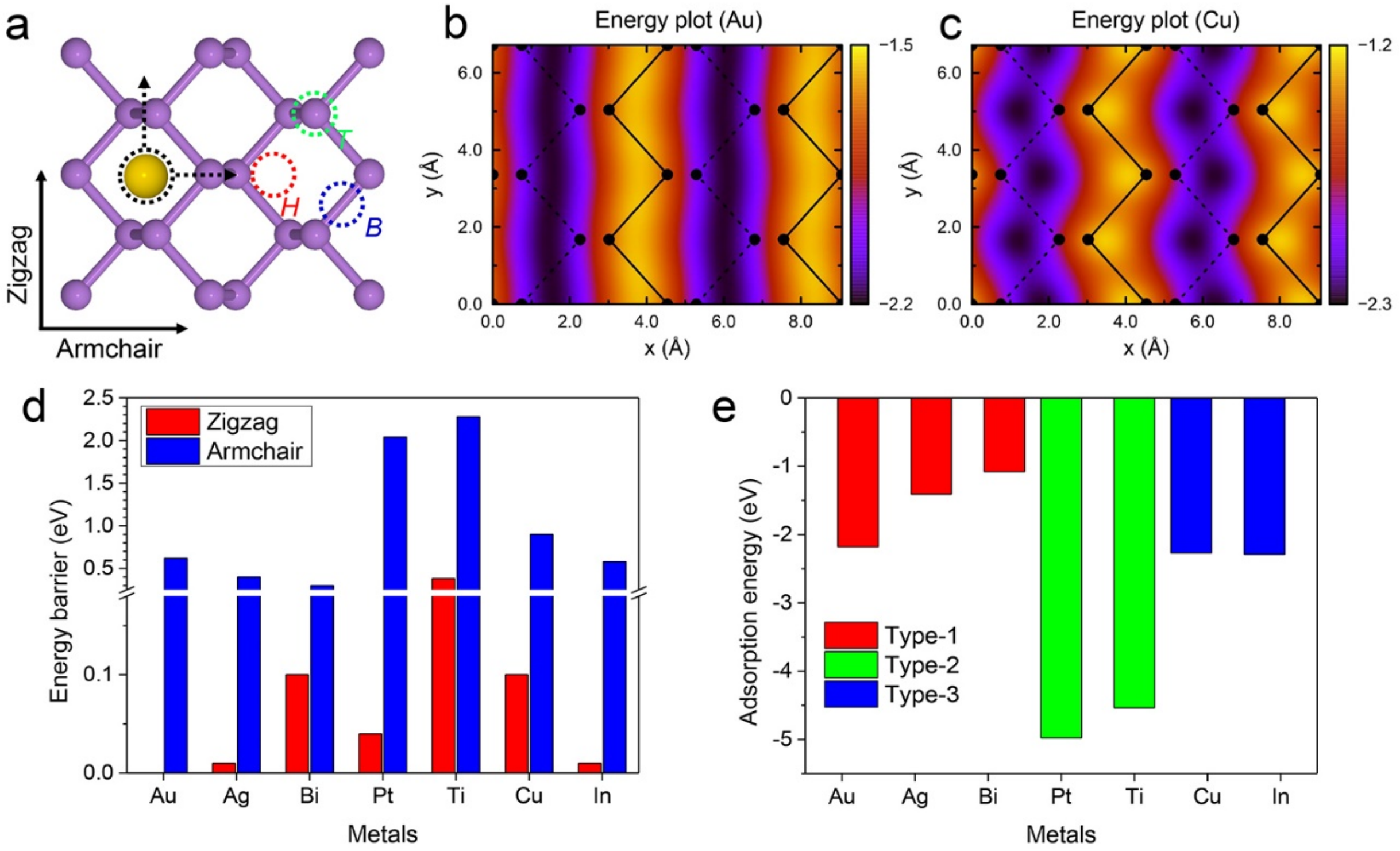

**Figure 5. Theoretical calculations about the interactions of metals and BP.** (a) Atomic model used for calculation. (b, c) Adsorption energy ($E_a$) plot for Au and Cu atoms on BP. (d) Energy barrier for adatom diffusion along the zigzag (red) and armchair (blue) directions on BP. (e) Adsorption energy of metal adatoms on BP at hollow (*H*) sites. Color denotes the different growth modes (red: single-crystalline growth, green: poly-crystalline growth, blue: metal phosphide formation).

ASSOCIATED CONTENT

**Supporting Information**. The Supporting Information is available free of charge on the ACS Publications website.

Schematic of metal/BP sample fabrication process, AFM topography images, TEM/STEM characterization data of samples, interfacial binding energy calculation and structure relaxation of Au on BP, extra XPS spectra of Au-BP, summarized the lattice mismatch between BP and investigated metals, and calculated Gibbs free energies of metal phosphides at 298 K.

**Corresponding Author**

E-mail: kpkim@yonsei.ac.kr, h.j.choi@yonsei.ac.kr,and hulex@unist.ac.kr

**Notes**

The authors declare no competing interests.

**Acknowledgements**

We are grateful to Yeonjin Yi and Hyunbok Lee for helpful discussion. This work was mainly supported by the Basic Science Research Program at the National Research Foundation of Korea (NRF-2017R1A5A1014862 and NRF-2019R1C1C1003643), by the Yonsei Signature Research Cluster Program of 2021 (2021-22-0004), and by the Institute for Basic Science (IBS-R026-D1). Y.L. received support from the Basic Science Research Program at the National Research Foundation of Korea which was funded by the Ministry of Education (NRF-2020R1A6A3A13060549), Ministry of Science and ICT (NRF-2021R1C1C2006785), and from the 2020 Yonsei University Graduate School Research Scholarship Grants. H.-g.K. and H.J.C. are supported by the NRF of Korea (Grant No. 2020R1A2C3013673). Computational resources have been provided by KISTI Supercomputing Center (Project No. KSC-2019-CRE-0195). H.Y.J. acknowledges the supports from the Basic Science Research Program at the National Research Foundation of Korea (NRF-2018R1A2B6008104). G.H.L. acknowledges the supports from the Korea Institute of Energy Technology Evaluation and Planning (KETEP) and the Ministry of Trade, Industry & Energy (20173010013340) and Creative-Pioneering Researchers Program through Seoul National University(SNU).